# Optimal Scheduling of Electric Vehicles Charging in low-Voltage Distribution Systems

Shaolun Xu\*, Liang Zhang\*\*, Zheng Yan†, Donghan Feng\*, Gang Wang\*\* and Xiaobo Zhao\*

**Abstract** – Uncoordinated charging of large-scale electric vehicles (EVs) will have a negative impact on the secure and economic operation of the power system, especially at the distribution level. Given that the charging load of EVs can be controlled to some extent, research on the optimal charging control of EVs has been extensively carried out. In this paper, two possible smart charging scenarios in China are studied: centralized optimal charging operated by an aggregator and decentralized optimal charging managed by individual users. Under the assumption that the aggregators and individual users only concern the economic benefits, new load peaks will arise under time of use (TOU) pricing which is extensively employed in China. To solve this problem, a simple incentive mechanism is proposed for centralized optimal charging while a rolling-update pricing scheme is devised for decentralized optimal charging. The original optimal charging models are modified to account for the developed schemes. Simulated tests corroborate the efficacy of optimal scheduling for charging EVs in various scenarios.

## 1. Nomenclature

The main notation used in the paper is presented below for quick reference. Other symbols are defined where needed.

| | |
|---|---|
| $B_c$ | Battery Capacity. |
| $c$ | Constant charging price for centralized optimal charging. |
| $C_t$ | Total charging revenues in time interval $t$. |
| $E_i^{soc}$ | The minimum desired state of charge of $EV_i$. |
| $J$ | The last plug-off time interval. |
| $L_{t,j}$ | Total load in time interval $j$ after time interval $t$. $L_{0,j}$ is the base load in time interval $j$. |
| $L_t^{p-v}$ | The peak-valley difference in time interval $t$, $L_0^{p-v}$ is peak-valley difference of the base load and $L_{96}^{p-v}$ is the peak-valley difference of the total load. |
| $M_i$ | Daily travel miles of $EV_i$. |
| $N$ | The number of plug-in EVs. |
| $p_j$ | Electricity prices in time interval $j$. |
| $p_{t,j}$ | Real-time price in time interval $j$ after time interval $t$ for decentralized optimal charging. |
| $P_{MTF}$ | Maximum load capacity of the distribution transformer. |
| $P_r$ | Rated Charging Power. |
| $R_i^{soc}$ | Real state of charge of $EV_i$. |
| $R_{p-v}$ | Incentives for the reduction of peak-valley difference. |
| $S_i^{soc}$ | Start state of charge (SOC) of $EV_i$. |
| $S_{i,j}$ | The control variable of $EV_i$ in time interval $j$, $S_{i,j} = \begin{cases} 1, & EV_i \text{ is charging in time interval } j \\ 0, & \text{otherwise} \end{cases}$ |
| $\Delta t$ | Length of a time interval. |
| $T_i^a$ | Plug-in time of $EV_i$. |
| $T_i^d$ | Plug-off time of $EV_i$. |
| $\eta_c$ | Charging efficiency. |

## 2. Introduction

As an emerging effective way to mitigate the $CO_2$ emissions and oil dependency, EVs have become the focus of the automotive industry and the government [1-2]. According to the EV development strategy research report by the Ministry of Industry and Information Technology of China, there will be about 60 million EVs in 2030 in China. Many cities in China, like Beijing and Shanghai have promulgated a series of policies to encourage the use of EVs.

With large numbers of EVs plugged-in, the overall load profile will be greatly affected. Uncoordinated charging of large-scale EVs will ineluctably have a negative influence on the secure and economic operation of power system, especially at the distribution level [3-5]. Supposing that the charging load of EVs can be controlled to some extent, different optimal charging methods have been proposed in the literature.

† Corresponding Author: Dept. of Electrical Engineering, Shanghai Jiao Tong University, China. (yanz@sjtu.edu.cn)
\* Dept. of Electrical Engineering, Shanghai Jiao Tong University, China.
\*\* Dept. of Electrical and Computer Engineering, University of Minnesota, America.

Centralized control strategies are developed in [6-10]. Random numbers are generated to characterize the EV plug-in time [6]. Yet the specific charging requests of individual EV users are neglected. The charging rate is optimized to maximize the total charging capacity within network limits [7]. An improved two-stage optimization model is proposed to increase the economic benefit of charging station and reduce the peak-valley difference [8]. An optimal charging scheduling scheme is reported to minimize the total charging cost under TOU price [9]. Nonetheless, new load peaks will arise in the valley price period. Global and local optimal scheduling schemes for charging of EVs are studied in [10]. In aforementioned optimal charging approaches, the number of control variables increase drastically with the number of EVs. To address the dimensional problem, decentralized control strategies receive more and more attention.

In [11-17], different decentralized control strategies are discussed. The scheduling of EV charging is optimized to maximize benefits of consumers [11-15]. Iterative decentralized optimization schemes based on Lagrange relaxation method are proposed in [11-13]. Decentralized mechanisms are devised based on congestion pricing that is used in IP networks, but the optimality cannot be guaranteed [14]. Distribution locational prices are leveraged to guide the charging of EVs [15]. This method does not require an iterative communication and computation process, but the dc optimal power flow model used in attaining locational price may cause trouble given that line resistances in distribution networks are relatively large. A decentralized algorithm is developed to iteratively solve the valley-filling problem with provable convergence to the optimality. But to ensure the optimality, an additional term is essential in the end user's response function, i.e., the end user's objective is no longer purely maximum economic benefits oriented. The additive increase and multiplicative decrease charging algorithm are enhanced to take local voltage constraints into account [16]. Yet in this scheme, the charging requests of end users (such as charging time requests) are overlooked and the charging process is still under the grid cooperation control.

Centralized and decentralized optimal charging of EVs are both considered in this paper. In centralized optimal charging, charging of EVs is managed by aggregators (EV charging station can also be viewed as a special aggregator), while in decentralized charging it is managed by individual users. Without loss of generality, we assume that the aggregators and individual EV users only care about charging revenues. Therefore, under the current TOU pricing framework, a new load peak will occur at the very beginning of the valley price period 0. With high penetration of EVs, this new load peak can be even higher than the original one.

The present work targets the aforementioned new load peak problem, which can happen in the TOU pricing based optimal charging. The contributions of this paper are twofold: 1) A simple peak-valley difference related incentive mechanism is proposed for centralized charging management. This mechanism is free of bi-directional communication and complex computation; 2) A rolling-update pricing mechanism is devised for decentralized charging management. The proposed method only requires solving the optimization once while the existing decentralized valley-filling one requires iterative computation for both gird coordinator and individual users [17].

The remaining sections are organized as follows. Section 3 presents the centralized and decentralized charging architectures. Centralized optimal charging is detailed in Section 4 while decentralized optimal charging is specified in Section 5. Section 6 shows simulation results in different charging scenarios. Finally, Section 7 draws the conclusion.

## 3. Optimal Charging Control Architectures

In general, two optimal charging control architectures can be deployed, i.e., centralized and decentralized control 0. The difference between the two architectures mainly lies in locations where optimization decisions are made. In centralized control, there is a control center at the aggregator level, and in decentralized control, the optimal scheduling is performed by individual EV users.

The two architectures also differ with respect to computational complexity, implementation flexibility and information exchange requirements. In this paper, the two charging control architectures are detailed in section 4 and section 5, respectively.

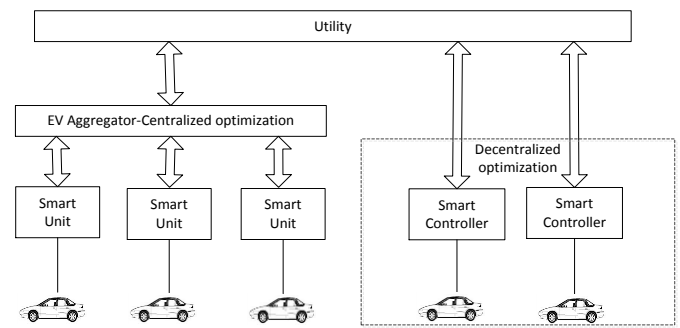

**Fig. 1.** EV charging architecture

The charging scenario considered in this paper is set in the low-voltage distribution network. The following customary assumptions are made: 1) For centralized optimal scheduling, charging spots of EVs under one distribution transformer are owned by one aggregator. The aggregator can get the profile of the base load day-ahead. 2) The spatial variation of the electricity price is neglected, which means the electricity price is the same at all locations at a given time instant. 3) The charging control method is the on-off method while the lithium-ion battery charging characteristic is applied to simulate the charging behavior.

## 4. Centralized Optimal Charging

Centralized optimal scheduling: The EV aggregator collects the charging requests of EV users to make the optimal charging schedules.

### 4.1 Charging Scenario Description

Figure 2 illustrates the centralized optimal charging scenario.

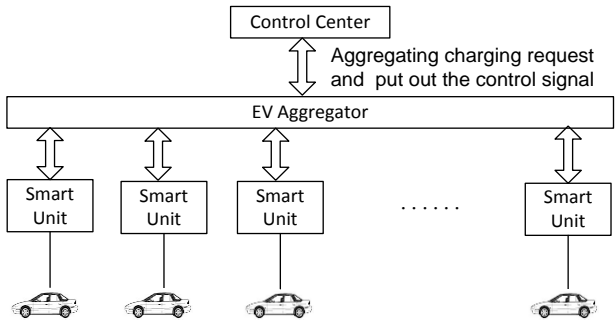

**Fig. 2.** Centralized optimal charging scenario

When an EV is connected to the charger and requests charging, the smart unit uploads the current state of charge (SOC), battery type and total capacity of the battery to the aggregator. Meanwhile, the EV user should set the plug-off time and the expected SOC. Under the premise of satisfying the charging request of EVs, the aggregator determines the optimal charging schedule to maximize its total benefits.

The ideal solution requires perfect knowledge of EV plug-in, plug-out, SOC information and also base load over the scheduling period, which is infeasible in practice. Akin to the local optimal scheduling in 0, our method divides one day into 96 optimization time intervals, and calculates the optimal charging schedule at the end of each interval for EVs arriving in that interval.

### 4.2 Centralized Optimal Scheduling Formulation

In time interval $t$, the scheduling period is from $t$ to $J$ which is the last plug-out time interval among the $N_t$ EVs:

$$J = |\max_{i \cdots N_t}(T_i^d) / \Delta t|. \quad (1)$$

The total charging revenues of the aggregator are

$$C_t = \sum_{i=1}^{N_t} \sum_{j=t+1}^{J} P_r S_{i,j} \Delta t (c - p_j). \quad (2)$$

The charging price $c$ should be set to be constant since the charging process is under the control of the aggregator.

The aggregator has a tendency to maximize its benefits, i.e., the charging revenues (2) in this situation. In the optimal scheduling process, the following constraints should be considered.

1) The charging request of EV users. At the plug-out time, the SOC of $EV_i$ should be no less than the expected SOC and should be no greater than one:

$$S_i^{soc} B_c + \sum_{j=t}^{J} P_r \eta_c S_{i,j} \Delta t \geq E_i^{soc}, \quad i = 1, \ldots, N_t \quad (3)$$

$$S_i^{soc} B_c + \sum_{j=t}^{J} P_r \eta_c S_{i,j} \Delta t \leq B_c, \quad i = 1, \ldots, N_t. \quad (4)$$

Let $J_i^d = |T_i^d / \Delta t|$. In the time interval after $J_i^d$, $EV_i$ should be in the non-charging state, i.e.

$$S_{i,j} = 0, \cdots j = J_i^d + 1, \ldots, J, \quad i = 1, \ldots, N_t. \quad (5)$$

2) The transformer capacity constraint. For the sake of security, in time interval $t$, the total load should be less than the maximum loading capacity of the distribution transformer:

$$L_{t-1,j} + \sum_{i=1}^{N_t} P_r S_{i,j} < P_{MTF}, \quad j = t, \ldots, J. \quad (6)$$

After time interval $t-1$, the total load in time interval $j$, $L_{t-1,j}$ is

$$L_{t,j} = L_{t-1,j} + \sum_{i=1}^{N_t} P_r S_{i,j}, \quad j = t, \ldots, J. \quad (7)$$

The problem defined by (2)-(7) is linear optimization with $S_{i,j}$ as control variables.

### 4.3 Optimal Scheduling Considering Grid Incentives

Time of use (TOU) electricity price mechanism is utilized extensively in China. Centralized optimal scheduling with TOU price will result in new load peaks in the valley price period 0, which is shown in Fig. 3.

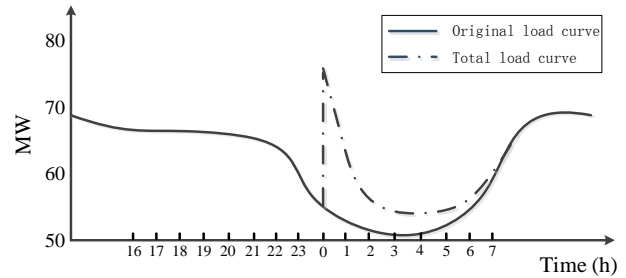

**Fig. 3.** Centralized optimal charging with TOU price

As clearly seen in Fig. 3, a new load peak arises. The reason is that the majority of EVs plug-in in the evening 0 and are arranged to charge when the electricity price is low. With the increasing number of EVs, the new load peak in the valley period will become very high (even higher than the original load peak). The new load peak requires a severe high ramping rate, increases ramping cost, endangers power system security, and reduces power system reliability. In the near future, this new load peak will inevitably impede high penetration of EVs.

To mitigate the potential new load peak, in contrast to the iterative incentive scheme in 0, we propose that the Grid Corporation should give incentives to the EV aggregators for their efforts in reducing peak-valley difference.

Assuming a nonnegative incentive factor $\alpha$, the total incentives given to the aggregator can be set as follows:

$$R_{p-v} = \begin{cases} \alpha(L_0^{p-v} - L_{96}^{p-v}), & L_{96}^{p-v} \leq L_0^{p-v} \\ 0, & L_{96}^{p-v} \geq L_0^{p-v} \end{cases} \quad (8)$$

where $L_t^{p-v} = \max_j(L_{t,j}) - \min_j(L_{t,j})$ and $\alpha$ represents the Grid Corporation's willingness to pay for the reduction in peak-valley difference.

In the sequel, the aggregator's optimization model needs to be modified. With grid incentives, the aggregator's total benefits after one day operation become:

$$F = \sum_{t=1}^{96} C_t + R_{p-v} \quad (9)$$

To maximize the benefits, the aggregator needs to maximize its charging revenues while minimizing the peak-valley difference of the total load.

Due to the lack of future arrival information of EVs, it is impossible to get the quantitative relation between the global peak-valley difference and the current peak-valley difference. In the local time interval, the aggregator cannot calculate the benefits for its efforts in reducing the peak-valley difference. Based on a greedy algorithm, the aggregator can minimize the local peak-valley difference to obtain the approximate minimum global peak-valley difference. Thus, in the local time interval $t$, the aggregator aims to maximize the local charging revenues while minimize the local peak-valley difference. For the aggregator, the optimal scheduling gives rise to a multiple-objective optimization. To formulate such a problem, the weighted coefficient method can be used.

In the time interval $t$, the objective of the aggregator can be written as:

$$\max \quad C_t - \beta L_t^{p-v} \quad (10)$$

where $\beta$ is a weighting coefficient, dictating the importance of the second part.

One possible way to set $\beta$ is to make it small enough to ensure Pareto optimality, which guarantees the maximum charging revenues to be obtained. Though this setting may cause deficiency in the overall benefits to the aggregator, it can make sure that the overall benefits are no less than the benefits obtained without grid incentives. One possible setting of $\beta$ can be

$$\beta = 10^{-5} / L_0^{p-v}. \quad (11)$$

As presented in Section *B*, the constraints can be formulated as (3)-(7).

## 5. Decentralized Optimal Charging

In decentralized optimal scheduling, each EV owner is postulated to have a smart controller which can receive the price signal issued by the aggregator or the utility. The smart controller performs local optimal scheduling.

### 5.1 Charging Scenario Description

Figure 4 depicts the decentralized optimal charging scenario. When an EV is connected to the charger, the smart controller collects battery information. At the same time, the EV user is required to set its plug-out time and minimum expected SOC. With the received charging price, each individual smart controller schedules the EV charging to minimize the charging cost.

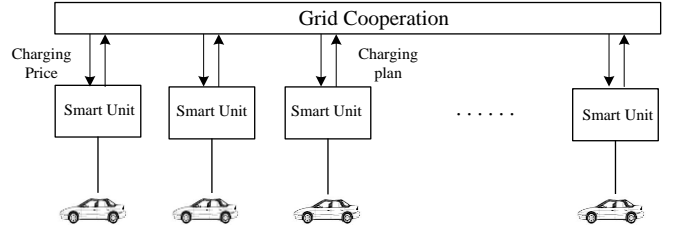

**Fig. 4.** Decentralized optimal charging scenario

### 5.2 TOU Price Guided Decentralized Optimal Scheduling

In TOU pricing, the charging price is fixed. For an individual EV, the optimization model can be formulated as follows:

$$\min \sum_{j=t+1}^{J_i} P_r S_{i,j} \Delta t p_j \quad (12)$$

$$E_i^{SOC} B_c \leq S_i^{SOC} B_c + \sum_{j=t}^{J_i} P_r \eta_c S_{i,j} \Delta t \leq B_c \quad (13)$$

This linear optimization model can be easily solved.

### 5.3 Rolling-update Price Guided Decentralized Optimal Scheduling

TOU pricing guided decentralized optimal scheduling can give rise to a similar problem shown in Fig. 3. To mitigate the peak-valley difference, a new pricing is devised for decentralized optimal EV charging. Dividing one day into 96 time intervals and after each time interval, the pricing is updated as

$$p_{t,j} = aL_{t,j} / P_{MTF} + b, \quad j = 1 \cdots 96 \quad (14)$$

where $a$ is the slope, indicating the changing rate of the rolling-update price with respect to the total load; $b$ is the intercept, representing the price when there is no load. Both $a$ and $b$ are pre-determined factors and $a$ is required to be non-negative. It is worth mentioning that the new pricing will not contradict the normal locational marginal prices (LMPs) 0. Here, normal indicts multiple 0 and non-convex 0 issues of LMPs are overlooked. LMPs only take the transmission network into consideration where the distribution network is treated as one load node. The proposed pricing is designed for demand management in the distribution network especially for EV charging.

The implementation of the pricing is detailed in Fig.5

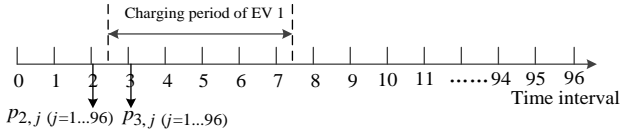

**Fig. 5.** Rolling-update price

The new pricing proposed in this paper will be referred to as rolling-update pricing. The protocol of this pricing is:

**Protocol** for rolling-update charging price:

i) For EVs plugging-in in time interval *t*, each smart controller determines the optimal charging plan according to the charging price announced after time interval *t-1*.

ii) The optimal charging plans are required to be uploaded to the utility. At the end of time interval *t*, the total load in time interval *j* will be updated as (7).

iii) The charging price for EVs plugging-in in time interval *t+1* will be updated according to (14).

Taking Fig.5 as an example, $EV_1$ plugs-in in time interval 3. The charging price for $EV_1$ is $p_{2,j}$, and the charging cost for $EV_1$ is $\sum_{j=t+1}^{J} P_r S_{i,j} \Delta t p_{2,j}$. The optimal charging plan for $EV_1$ will be uploaded to the utility. At the end of time interval 3, the load and the charging price will be updated.

With the rolling-update charging price, the objective function of individual EV users will be:

$$\min \sum_{j=t+1}^{J} P_r S_{i,j} \Delta t p_{t,j} \quad (15)$$

The constraint is (13).

The rolling-update price can reflect the dynamic change of the total load. In the time interval with high load level, the charging price will be high. This characteristic of the rolling-update charging price is favorable to reducing the peak-valley difference.

## 6. Case Studies

The centralized and decentralized charging scheduling algorithms developed in Section IV and Section V are simulated here. Drawbacks of TOU pricing based optimal charging are also depicted in this section.

Basic simulation settings in this paper include

i) Battery capacity: 32 kW·h;

ii) Rated charging power: 7 kW, charging efficiency: 90%;

iii) Based on the maximum likelihood estimation method 0 and the original travel survey data 0, the plug-in time, plug-out time and daily travel miles probability density functions can be respectively defined as (16), (17) and (20):

$$f_s(x) = \begin{cases} \frac{1}{\sqrt{2\pi}\sigma_S} \exp[-\frac{(x+24-\mu_S)^2}{2\sigma_S^2}], \\ \qquad 0 < x \leq \mu_S - 12 \\ \frac{1}{\sqrt{2\pi}\sigma_S} \exp[-\frac{(x-\mu_S)^2}{2\sigma_S^2}], \\ \qquad \mu_S - 12 < x \leq 24 \end{cases} \quad (16)$$

where $\mu_s = 17.47$; $\sigma_s = 3.41$;

$$f_e(x) = \begin{cases} \frac{1}{\sqrt{2\pi}\sigma_e} \exp[-\frac{(x-\mu_e)^2}{2\sigma_e^2}], \\ \qquad 0 < x \leq \mu_e + 12 \\ \frac{1}{\sqrt{2\pi}\sigma_e} \exp[-\frac{(x-24-\mu_e)^2}{2\sigma_e^2}], \\ \qquad \mu_e + 12 < x \leq 24 \end{cases} \quad (17)$$

where $\mu_e = 8.92$; $\sigma_e = 3.24$;

$$f_m(x) = \frac{1}{\sqrt{2\pi}\sigma_m x} \exp[-\frac{(\ln x - \mu_m)^2}{2\sigma_m^2}] \quad (18)$$

where $\mu_m = 2.98$; $\sigma_m = 1.14$.

The plug-in, plug-out time and daily travel miles of each EV are generated from probability distribution (16)-(18).

iv) The expected SOC of the user is set to be 90%;

v) Energy needed per 100KM: $E_{100}$=15 kW·h. The start SOC of the battery can be calculated by:

$$S_i^{soc} = R_i^{soc} - M_i E_{100} / (100 B_c) \quad (19)$$

vi) The charging scenario is set in a residential area as shown in Fig. 6.

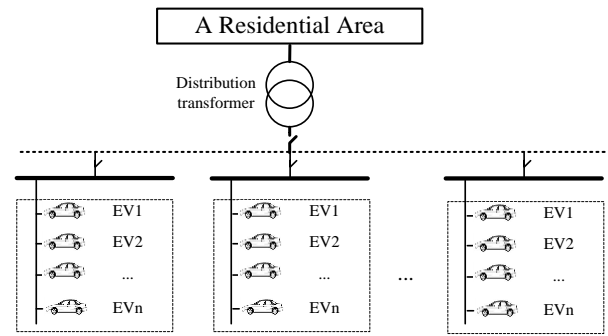

**Fig. 6.** Charging scenario in a residential area

The rated capacity of the distribution transformer is 6300 kVA. Assuming that power factor is 0.85 and energy efficiency is 0.95, the maximum loading capacity of the transformer is:

$$P_{MTF} = 6300 \cdot 0.85 \cdot 0.95 = 5087 \text{ kW} \quad (20)$$

The base load curve is set according to the typical residential load curve in Shanghai, China.

vii) The TOU price is set as the industrial electricity price in China, which is shown in table 1 0.

**Table 1.** TOU electricity price

| Period | Electricity price/ (Yuan/kW·h) |
|---|---|
| Valley period (00:00—08:00) | 0.365 |
| Peak period (08:00—12:00,17:00—21:00) | 0.869 |
| Level period (12:00—17:00, 21:00—00:00) | 0.687 |

Charging of 150 and 300 EVs is simulated on MATLAB+CPLEX. The main function is programmed in MATLAB while CPLEX is used for solving the optimization. The simulation horizon is from 12:00 pm to 12:00 pm next day, i.e., the time slot 0 in Fig. 5 represents 12:00 pm.

### 6.1 Uncoordinated Charging

1) Uncoordinated charging: All EVs begin charging as soon as they plug-in, and stop charging when the battery is full. The simulation results are shown in Fig. 7, which clearly indicate that uncoordinated charging of a large number of EVs will increase the load peak tremendously. The maximum loading capacity of the transformer is 5087kW and uncoordinated charging of 150 EVs will endanger the secure operation of the transformer.

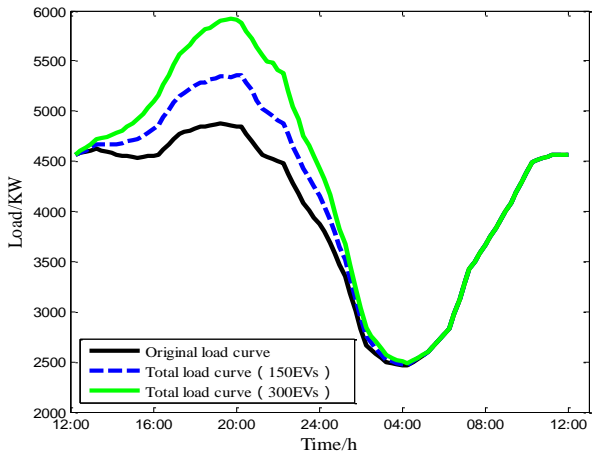

**Fig. 7.** Load curves after uncoordinated charging of 150 and 300 EVs

### 6.2 Centralized Optimal Charging

1) Without grid incentives for peak-valley difference reduction, the aggregators schedule the charging of EVs to maximize the charging revenues. The resultant load curves are presented in Fig. 8, from which we can see that all EVs are scheduled to charge in the valley-price period. Yet new load peaks occur. When charging 300 EVs, the new load peak becomes higher than the original one.

2) Set incentive factor $\alpha = 0.1$ Yuan/kW. The original peak-valley difference of base load is 2416.0kW, thus according to (11) setting $\beta = 4 \times 10^{-9}$. With the grid incentives, the aggregators schedule the charging of EVs to maximize their total benefits. The resultant load curves are shown in Fig. 9. In this case, the shape of the total load curves becomes smooth and the new load peak is much lower.

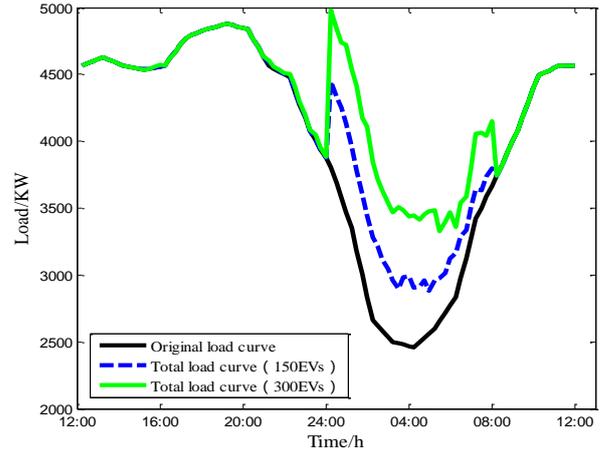

**Fig. 8.** Total load curves after centralized optimal charging 150 and 300 EVs without Grid incentives

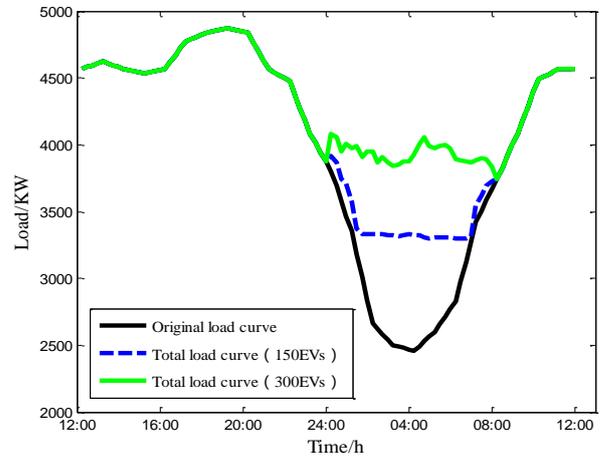

**Fig. 9.** Total load curves after centralized optimal charging 150 and 300 EVs with Grid incentives

3) Economic benefits of the aggregator and the peak-valley difference are compared in the following charging scenarios.

**Table 2.** Charging revenues in uncoordinated charging and centralized optimal charging

| EV Number | Charging revenues /Yuan | | |
|---|---|---|---|
| | Uncoordinated | Without incentives | With incentives |
| 150 | 1113.2 | 2218.5 | 2218.5 |
| 300 | 2314.6 | 4605.3 | 1605.3 |

**Table 3.** Total benefits in uncoordinated charging and centralized optimal charging

| EV Number | Total benefits/Yuan | | |
|---|---|---|---|
| | Uncoordinated | Without incentives | With incentives |
| 150 | 1113.2 | 2218.5 | 2262.3 |
| 300 | 2314.6 | 4605.3 | 4648.5 |

**Table 4.** Peak-valley difference comparison in uncoordinated charging and centralized optimal charging

| EV Number | Peak-valley difference/kW | | |
|---|---|---|---|
| | Uncoordinated | Without incentives | With incentives |
| 150 | 2892.4 | 2014.7 | 1576.8 |
| 300 | 3441.7 | 1561.2 | 1129.5 |

Original peak-valley difference in base load is 2416.0kW. For centralized optimal charging of 150 and 300 EVs, the peak-valley difference can be further reduced by 437.9kW and 431.7kW, respectively, when the incentives are offered by grid. The offered incentives are 43.8Yuan and 43.2Yuan, respectively.

Simulation results show the effectiveness of the proposed incentive scheme in further reducing the peak-valley difference and smoothing the load curve. The cost of this incentive is trivial to the grid. In addition, the aggregators will support the incentive mechanism since they can improve the total benefits with little effort in updating their optimization model.

In a nutshell, the simple incentive mechanism proposed in this paper can help avoid new load peak and smooth the load curve, which overcomes drawbacks in TOU pricing.

### 6.3 Decentralized Optimal Charging

1) TOU pricing guided decentralized optimization. The simulation results with TOU price are presented in Fig. 10.

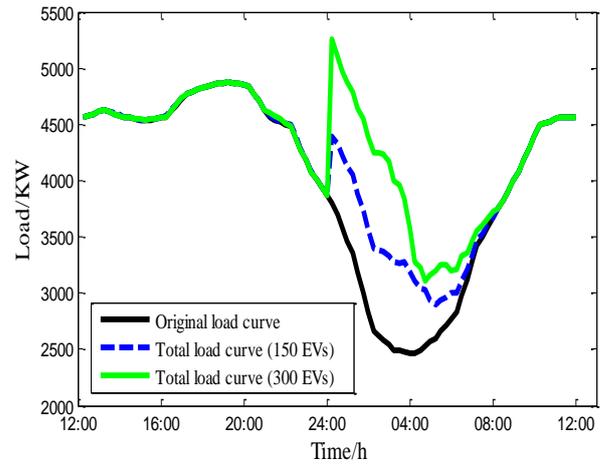

**Fig. 10.** Total load curves after decentralized optimal charging of 150 and 300 EVs with TOU price

Figure 10 indicates that all EVs are controlled to charge in the valley price period and a new load peak will arise in that period. The arising of new load peak is due to the relatively static state of TOU pricing. Pricing method such as TOU, hourly pricing and real time pricing cannot reflect the EV load in a dynamic manner, which leads to the new peak load.

2) Rolling-update price guided decentralized optimal scheduling. The parameters of the rolling-update price are set as follows: $b$ is set to be zero; $a$ is set according to the original load profile and the TOU price. In the original load profile, the maximum load in the valley price period is 3803.7kW. Setting $a*0.9*3803.7/5087=0.36$ gives $a=0.542$. Simulation results with the rolling-update price are depicted in Fig. 11.

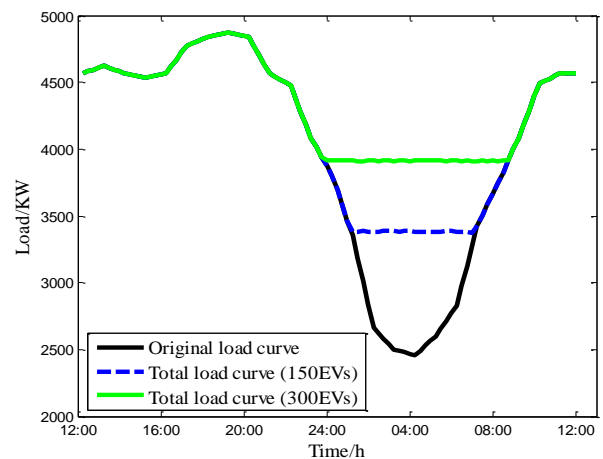

**Fig. 11.** Total load curves after decentralized optimal charging of 150 and 300 EVs with rolling-update price

In Fig.11. all EVs are scheduled to charge in the low price period. However, the shape of the resultant load curves becomes smooth and no new load peak arises because the rolling-update price can reflect the dynamic change of the total load.
3) The peak-valley difference is compared in different charging scenarios.

**Table 5.** Peak-valley difference comparison in uncoordinated charging and decentralized optimal charging

| EV Number | Peak-valley difference/kW |  |  |
|---|---|---|---|
|  | Uncoordinated | TOU price | Rolling-update price |
| 150 | 2892.4 | 2199.0 | 1576.8 |
| 300 | 3441.7 | 2303.1 | 1077.3 |

From table 5, when rolling-update price instead of the TOU price is used as the charging price, the peak-valley difference of the total load can be significantly reduced. The simulation results corroborate the efficacy of rolling-update price mechanism.

Comparing the effectiveness of the incentive mechanism and the rolling-update price, when the number of EVs are relatively small (150 EVs in this case), the two mechanisms can achieve similar results. Under the assumption that the aggregators tend to maximize the charging revenue first, the rolling-update price is much more effective in reducing the peak-valley difference when the number of EVs is relatively large.

## 7. Conclusion

The uncoordinated charging of EVs will increase the load peak tremendously since the using habits of EV owners. To mitigate the negative influence, this paper proposes two charging scenarios of optimal scheduling in China, which are centralized optimal charging deployed by the aggregator and decentralized optimal charging managed by individual users. However, with TOU price, maximizing the benefits of the aggregator and minimizing the cost of individual users will cause new peak loads in the valley price. Hence, this paper devises two scheduling methods which can be correspondingly applied to the centralized optimal charging by an aggregator and decentralized optimal charging managed by individual users. Ones is giving the aggregator the economic incentive with respect to the peak-valley difference and the other one is the rolling-update pricing method for individual users. The simulation results corroborate that for centralized management scenario, the proposed incentive mechanism is effective in further reducing the peak-valley difference and smoothing load curve; for decentralized management, the developed rolling-update pricing can achieve desirable efficiency in valley-filling. Further study will be focused on the incentive mechanism or pricing method taking V2G into consideration.

## Acknowledgements

The authors would like to thank the State Energy Smart Grid R&D Center and Alstom Grid-SJTU Joint Research Center, Shanghai, China, for its support.

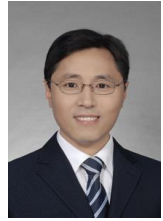

**Shaolun Xu** He received his M.S in Department of Electrical Engineering, North China Electric Power University in 2004 and now pursues the Ph.D. degree in Shanghai Jiao Tong University (SJTU). His research interests are cyber physical system and coordinative charging of electric vehicles in smart grids.

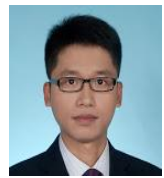

**Liang Zhang** He received his diploma, B.Sc. and M.Sc. in Electrical Engineering from Shanghai Jiao Tong University in 2012 and 2014, respectively. Since then, he has been working toward the Ph.D. degree in the ECE Dept. of the Univ. of Minnesota. His current research interests span the areas of optimization, monitoring, and inference for power systems.

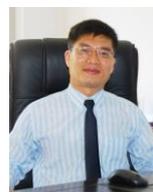

**Zheng Yan** He received his B.S, M.S and Ph.D. degrees, all in Electrical Engineering, from Shanghai Jiao Tong University in 1984, and Tsinghua University in 1987 and 1991, respectively. Before joining EE Dept. of Shanghai Jiao Tong University in 2004 as a full Professor, he had been on research cooperation at Ibaraki University in Japan (1994-1996), Cornell University (1997-2001), the University of Hong Kong (2001-2004). His research interests are in application of optimization theory to power systems and power markets, and dynamic security assessment.

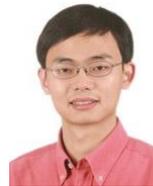

**Donghan Feng** He has been with the faculty of Shanghai Jiao Tong University since 2008. Currently, he holds the position of Associate Professor. He also serves as the Deputy Director of the State Energy Smart Grid R&D Center. Dr. Feng received his B.Sc. and Ph.D. degree both in the Department of Electrical Engineering, Zhejiang University in 2003 and 2008, respectively. He worked at Tsinghua University as a graduate research assistant during 2005-2006 and worked at University of Hong Kong as a visiting scholar during 2006-2007. Since 2009, he has been granted the Hans Christened Ørsted Funding for collaborative research with the Centre for Electric Technology, Technical University of Denmark. His research interests are market operation of smart grid and smart energy networks.

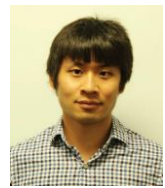

**Gang Wang** He received the B.Eng. degree in Electrical Engineering and Automation from Beijing Institute of Technology, Beijing, China in 2011. He is currently working toward his Ph.D. degree in the ECE Dept. of the Univ. of Minnesota. His research interests focuses on the areas of signal processing and optimization for smart power grids.

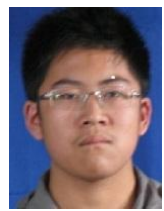

**Xiaobo Zhao** He received the B.Sc. Degree in Electrical Engineering, from Shanghai Jiao Tong University, Shanghai, China, in 2014. He is currently working toward the M.S degree in Electrical Engineering at the Shanghai Jiao Tong University. His research interests cover the areas of electricity markets and coordinative charging of electric vehicles in smart grids.